\documentclass[journal]{IEEEtran}
\usepackage{mathtools}
\usepackage{float}
\usepackage{subfigure}
\usepackage{multicol}
\usepackage{paralist}
\usepackage{bm}
\usepackage{amsfonts}
\usepackage{url}
\usepackage{multirow}
\usepackage{color}
\usepackage{booktabs}
\makeatletter
\newif\if@restonecol
\makeatother

\usepackage[linesnumbered,ruled,vlined]{algorithm2e}
\usepackage{algpseudocode}
\usepackage{amsmath}
\usepackage{cleveref}
\usepackage{amssymb}
\usepackage{graphicx}
\usepackage{epstopdf}

\usepackage{color}

\ifCLASSOPTIONcompsoc

\usepackage[nocompress]{cite}
\else
\usepackage{cite}
\fi

\ifCLASSINFOpdf

\else

\fi

\begin{document}
	\title{Toward Structural Controllability and Predictability in Directed Networks}
	
\author{Fei Jing,
        Chuang Liu,
        Jian-Liang Wu,
        and~Zi-Ke~Zhang~\IEEEmembership{Member,~IEEE,}
\IEEEcompsocitemizethanks{\IEEEcompsocthanksitem  F. Jing and J.-L. Wu are with Department of Mathematics, Shandong University, Jinan, 250100, China (e-mail: fei.jing@mail.sdu.edu.cn (F. Jing) and jlwu@sdu.edu.cn (J.-L. Wu)).
\IEEEcompsocthanksitem C. Liu is with Alibaba Research Center for Complexity Sciences, Hangzhou Normal University, Hangzhou, 311121, China (e-mail: liuchuang@hznu.edu.cn). 
\IEEEcompsocthanksitem Z.-K. Zhang is with College of Media and International Culture, Zhejiang University, and Center for Zhejiang Digital Development and Governance, Hangzhou 310028, China (email: zkz@zju.edu.cn, corresponding author).}
\thanks{Manuscript received xxxx }}

\IEEEtitleabstractindextext{
	\begin{abstract}
The lack of studying the complex organization of directed network usually limits to the understanding of underlying relationship between network structures and functions.
Structural controllability and structural predictability, two seemingly unrelated subjects, are revealed in this paper to be both highly dependent on the critical links previously thought to only be able to influence the number of driver nodes in controllable directed networks. Here, we show that critical links can not only contribute to structural controllability, but they can also have a significant impact on the structural predictability of networks, suggesting the universal pattern of structural reciprocity in directed networks. In addition, it is shown that the fraction and location of critical links have a strong influence on the performance of prediction algorithms. Moreover, these empirical results are interpreted by introducing the link centrality based on corresponding line graphs. This work bridges the gap between the two independent research fields, and it provides indications of developing advanced control strategies and prediction algorithms from a microscopic perspective.
	\end{abstract}
	
	}
	
	%
\maketitle	

\IEEEdisplaynontitleabstractindextext

\IEEEpeerreviewmaketitle
	
	\section{Introduction}\label{sec:introduction}
	\IEEEPARstart{I}{n} the overwhelming majority of network studies, structural controllability and predictability of networks are generally considered two separate issues. Structural controllability, a typical structure dependent problem, focuses on studying the strategy to find a subset of nodes or components to influence the evolutionary process in state space control~\cite{lin1974structural}. There is a vast class of methods has been proposed and verified to effectively control networks, such as finding the minimum number of driver nodes via maximum matching~\cite{lin1974structural,liu2011controllability,gao2014target,yan2017network}, and the Popov-Belevitch-Hautus (PBH) rank condition~\cite{yuan2013exact}.  Different from structural controllability as an application of network topology, structural predictability tends to seek the upper bound of restoring the original or future patterns for a given network topology through mining the potential likelihood of the existence of a link between two arbitrary nodes~\cite{guimera2009missing,lu2015toward,sun2020revealing,kitsak2020link,LIAO2019182,8876661,articleZhang}. Generally, if a \emph{perfect} prediction method does exist, it should be able to foresee the future evolutionary network structure, and hence, it can promisingly capture the underlying mechanism. Therefore, it has been widely applied to various fields, ranging from biological~\cite{barzel2013network,yu2008high,stumpf2008estimating,chen2020large} and world-trade~\cite{kosowska2020network} to social~\cite{liben2007link,wang2011human,leskovec2010predicting,butun2019predicting} and online recommender systems~\cite{zhou2010solving}.

Although both structural controllability and predictability have been well studied separately, there has been a lack of attention paid to the relationship between the two issues. The former mainly addresses finding driver nodes or critical links of directed networks, while the latter focuses on understanding to what extent the given network can be recovered at large. On the one hand, methods on structural predictability can lead to more complete yet substantial structure, and hence can be more reliable in finding the target node sets to control the given network. On the other hand, results of structural controllability can also inform the design of more effective network prediction algorithms.

\begin{figure}[!t]
    \centering
    \includegraphics[width=8.5cm,height=10.5cm]{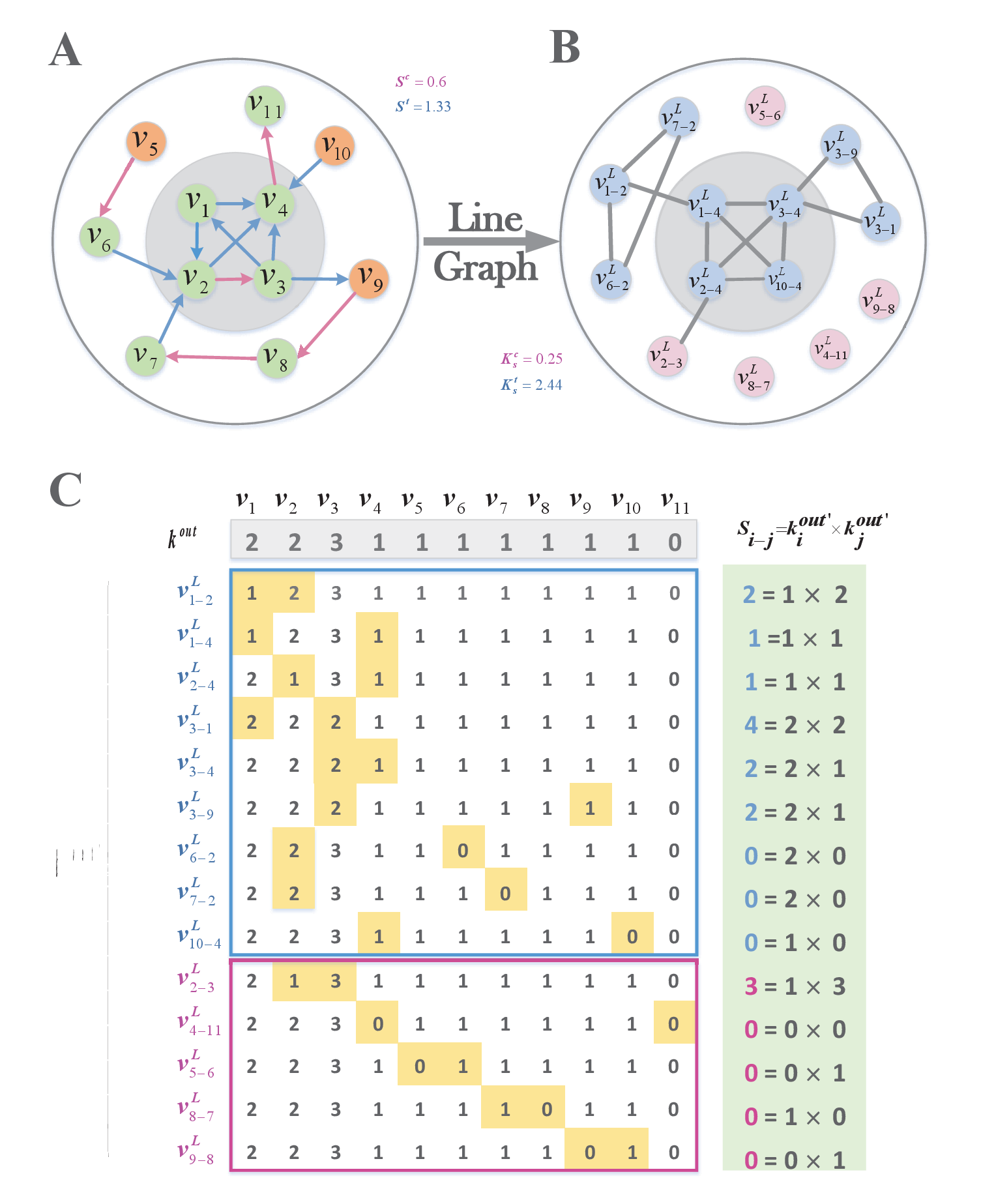}
    \caption{(A) Illustration of how critical and trivial links distribute in a directed network $G(V,V^L)$, where $V$ and $V^L$ respectively represent the node and edge set, and $V^L$ also represents the nodes of the corresponding line graph $G^{L}$. Pink and blue links represent critical and trivial links, respectively. Orange circles represent driver nodes. (B) The corresponding line graph ($G^{L}$) of network $G$. The transformed nodes of corresponding critical links are marked pink, e.g. $v^{L}_{2\!-\!3}$ corresponds to the critical link $V^L_{2\!-\!3}$ in $G$ and the edge between $v^{L}_{2\!-\!3}$ and $v^{L}_{2\!-\!4}$ have at least one common source node $v_2$. (C) Illustration of leave-one-out prediction process based on PA \cite{barabasi1999emergence} method for $G$. Each line and row represent the links and nodes in the original network $G$, and each number represents the excess out-degree ($k^{out'}$) once the corresponding link is removed, e.g., the original outdegree of $v_1$ is 2, and its excess outdegree of is 1 after removing the link $V^L_{1\!-\!2}$. Gray shaded numbers represent the original outdegree of each node ($k^{out'}$), and green shaded numbers show the calculation process, e.g., the PA scores for critical links $v^{L}_{8\!-\!7}$ and $v^{L}_{2\!-\!3}$ are $S_{8\!-\!7\!}\!=\!k^{out'}_8 \!\times \!k^{out'}_7\!=\!0$ and $S_{2\!-\!3\!}\!=\!k^{out'}_2 \!\times \! k^{out'}_3\!=\!3$, and for the trivial link  $S_{\!3\!-\!1\!}\!=\!k^{out'}_3\!\times\!k^{out'}_1=4$. Pink and blue numbers beside the graphs represent the average values of $S$ and $k_S$~\cite{kitsak2010identification} for critical and trivial links on $G$ and $G^{L}$, respectively.}
    \label{illustration}
\end{figure}

Here, we present a first step to reveal the relationship between structural controllability and predictability in directed networks from a microscopic perspective. The present observation begins by classifying network links into two categories according to the theory of structural controllability~\cite{jia2013emergence,posfai2013effect,liu2016control,maxwell1867,luenberger1964observing,zhu2019new,9376277,9354061}: 
(i) critical links, and (ii) trivial links. A link is defined as critical when the minimum size of driver nodes will increase if the very link is deleted (see \textbf{Methods}). Preliminary analyses show that critical links, in general, are more difficult to be identified and less predictable than trivial links. This is largely caused by the fact that most critical links distribute at the periphery of networks and share less information (Figure~\ref{illustration}a). On the contrary, there are also some exiguous circumstances that critical links locate in the network center (Figure~\ref{illustration}a), possessing rich information (e.g. more neighbors) and hence they can be easily predicted. As is shown in this paper, the structural predictability is highly dependent on the structural controllability, and therefore is called the ``structural reciprocity''.

Using the underlying concept of structural reciprocity, in this paper, the effects of critical and trivial links on the predictability of the entire network using representative algorithms are examined. Further analyses on network centrality features show that critical links play the intermediate role in connecting the structural controllability and  predictability in directed networks, and this metric can be used to distinguish the topological differences among various networks.\\

	\section{Theoretical Analysis}
	
	\label{sec:relatedwork}
	\subsection{Structural controllability}
Considering linear system dynamics based on a simple directed network~\cite{tanner2004leader,cheng2005advanced}, it is controllable~\cite{kalman1963mathematical,rahmani2009controllability} if and only if it can be driven from any initial state to any desired final state in finite time. Moreover, the network is called structurally controllable if it is possible to fix the free parameters in matrices $A$ and $B$, where $A$ and $B$ respectively describe the relationship among the internal parts of the linear control system and that between the internal and the controller, to certain values so that the obtained system ($A$,$B$) is controllable satisfying the Kalman rank condition~\cite{jia2013emergence, moore1981principal}. It has been proved that the system ($A$, $B$) is structurally controllable if and only if the minimum driver nodes set contains all of the unmatched nodes based on a maximum matching in $G$~\cite{lin1974structural,liu2011controllability}. A link is called a \emph{critical link} if the minimum number of driver nodes, which directly receive the external control signals \cite{poor2013introduction}, and if there is no directed link to any driver node in the maximum matching, will increase after removing it from the network; otherwise it is called \emph{trivial link}.  Therefore, the fraction of critical links is responsible for the robustness of structural controllability.\\

Given a directed network $G(V,E)$, we denote the in-degree ($p_k^{in}$) and  out-degree ($p_k^{out}$) distribution respectively as $\{p_0^{in}, p_1^{in},p_2^{in},\cdots\}$ and $p_k^{out}$ is $\{p_0^{out}, p_1^{out},p_2^{out},\cdots\}$. Consequently, for a sparse directed network, the number of driver nodes can be approximated as following \cite{L2006The,liu2011controllability}
\begin{align}
    n_D =& \frac{1}{2}\left\{[G(\hat{w}_2) + G(1-\hat{w}_1)-1] + [\hat{G}(w_2) + \hat{G}(1-w_1) \right. \nonumber \\
    & \left.-1] + \frac{z}{2}[\hat{w}_1(1-w_2) + w_1(1-\hat{w}_2)] \right\},
\end{align}
in which
\begin{align}
    G(x) &= \sum\limits_{i=0}^N p_i^{out} x^i \\
    \hat{G}(x) &=\sum\limits_{i=0}^N p_i^{in} x^i \\
    H(x) &= \sum\limits_{i=1}^N \frac{i p_i^{out}}{\langle k \rangle} x^{i-1} \\
    \hat{H}(x) &= \sum\limits_{i=1}^N \frac{i p_i^{in}}{\langle k \rangle} x^{i-1}
\end{align}
and $w_1,w_2,\hat{w}_1,\hat{w}_2$ satisfy the following self-consistent equations
\begin{align}
    w_1&=H(w_2) \\
    w_2 &= 1- H(1-\hat{w}_1) \\
    \hat{w}_1&=\hat{H}(w_2) \\
    \hat{w}_2 &= 1- \hat{H}(1-\hat{w}_1) .
\end{align}
	Next, we leave any link $(x,y)$ out from the original network, the in-degree sequence and out-degree sequence of the remaining network are labelled as  $\{\bar{p}_0^{in},\bar{p}_1^{in},\bar{p}_2^{in},\cdots\}$ and $\{\bar{p}_0^{out},\bar{p}_1^{out},\bar{p}_2^{out},\cdots\}$. Then, the number of driver nodes in the remaining network is
	\begin{align}
    n_D' =& \frac{1}{2}\left\{[G'(\hat{w}_2) + G'(1-\hat{w}_1)-1] + [\hat{G}'(w_2) + \hat{G}'(1-w_1) \right. \nonumber \\
    & \left.-1] + \frac{z}{2}[\hat{w}_1(1-w_2) + w_1(1-\hat{w}_2)] \right\},
    \end{align}
    in which
    \begin{align}
        G(x) &= \sum\limits_{i=0}^N \bar{p}_i^{out} x^i \\
        \hat{G}(x) &=\sum\limits_{i=0}^N \bar{p}_i^{in} x^i.
    \end{align}
    It is also obvious that
    \begin{align}
        \bar{p}_i^{out}=\left\{
             \begin{array}{ll} p_i^{out}-\frac{1}{N} & i=k_x \\
             p_i^{out}+\frac{1}{N} & i=k_x-1\\
             p_i^{out} &  otherwise
             \end{array}
        \right.
    \end{align}
    We define $\delta_D=n'_D-n_D$ as the difference between the number of driver nodes before and after the link $(x,y)$ is deleted.
    \begin{align}
        \delta_D &= \frac{1}{2N} \left\{ \hat{w}_2^{k_x^{out}}(\frac{1}{\hat{w}_2}-1)  + (1-\hat{w}_1)^{k_x^{out}}(\frac{1}{1-\hat{w}_1}-1) \right. \nonumber \\
        &\left. + w_2^{k_y^{in}}(\frac{1}{k_y^{in}} - 1) + (1-w_1)^{k_y^{in}}(\frac{1}{1-w_1}-1) \right\}
    \end{align}
    Naturally, it is known that $0\le \delta_D \le \frac{1}{N}$. And the link $(x,y)$ is critical if and only if $\delta_D=\frac{1}{N}$ according to the definition of critical links. It is worth mentioning that
    \begin{align}
        \frac{\partial \delta_D}{\partial k_x^{out}}=& \frac{1}{2N}\left\{\hat{w}_2^{k_x^{out}}\frac{(1-\hat{w}_2)\ln{\hat{w}_2}}{\hat{w}_2}  + \right. \nonumber \\ &\left. (1-\hat{w}_1)^{w_2}\frac{\hat{w}_1 \ln{(1-\hat{w}_1)}}{1-\hat{w}_1} \right\} \le 0
    \end{align}
    and
    \begin{align}
        \frac{\partial \delta_D}{\partial k_y^{in}}=&\frac{1}{2N}\left\{w_2^{k_y^{in}}\frac{(1-w_2)\ln{w_2}}{w_2}  + \right. \nonumber \\ &\left. (1-w_1)^{k_y^{in}}\frac{w_1 \ln{(1-w_1)}}{1-w_1} \right\} \le 0
    \end{align}
    It can be seen that, $\delta_D$ is a monotonically decreasing function of the two variables $k_x^{out}$ and $k_y^{in}$.
    This means that the link connecting two hub nodes is almost impossible to become a critical link. Conversely, the link connecting two marginal nodes is more likely to become a critical link. In the above proof, the tendency of critical links to be far away from the central area of the network is fully exposed. Although this proof is based on a sparse directed network with leaf nodes, we believe that this conclusion is still valid in real networks.

\subsection{Structural predictability}
Here, the leave-one-out approach~\cite{shirer2012decoding,wong2015performance} (see Figure~\ref{illustration}C) is adopted to measure the predictability of every target link. Given a simple directed network $G(V,E)$ consisting of $E$ and vacant set $\overline{E}$, only one particular link $e_{l}$ is deleted at a time from the original network $G$ and moved to a probe set $E^P$. And a selected prediction method is then used to score every unconnected node pair of $\overline{E}$ as well as $e_{ij}$. All of the pairs in $\overline{E}+e_{l}$ are then ranked in ascending order based on corresponding scores, and the normalized ranking of $e_{l}$ is used to quantify the predictability of that very link, which can be also obtained by a statistical sampling of comparisons~\cite{clauset2008hierarchical}. The higher the score, the larger the rank will be. Finally, the overall structural predictability of the network is finally obtained by averaging over all the links in $E$ using this approach.

Given a directed sparse network $G(V,E)$ with $N$ nodes and $M$ links. Obviously, we can have $M\ll \binom{N}{2}$. Before calculating the structural predictability of links, we can score every node pair $s_l$ in the original network with a given prediction method and rank them in increasing order. This sequence is denoted as $\{e_1,e_2,\cdots,e_{\mathcal{L}}\}$ satisfying that $s_1\le s_2\le \cdots \le s_{\mathcal{L}}$, where $1 \le l \le \mathcal{L}$ and $\mathcal{L}=\binom{N}{2}$. Generally, for any link prediction methods that are dependent on local topology, it will affect the scores of relatively a small number of node pairs, denoted as $\mathcal{O}{(N)}$, if we remove an arbitrary link $e_l$. 
Then, the resulting change in the normalized ranking number of $e_l$ in the sequence is
	\begin{equation}
	    \delta_l = \lim\limits_{N\to\infty}\frac{\mathcal{O}{(N)}}{\mathcal{L}} = 0.
	\end{equation}
That is to say, $P_l=\frac{l}{\mathcal{L}}$.
This suggests that, based on the leave-one-out approach, the structural predictability depends entirely on the topology of the original network before removing any single link. Taking PA method as an example, the node pair connecting two hub nodes tends to have greater structural predictability, while the node pairs far from the central area tend to have lower predictability. A similar conclusion is also true for the CN method. The node pairs in the core area often have more common neighbors and also have stronger predictability, while the node pairs far away from the central area are just the opposite. Although the above statement is strictly applicable to link prediction methods based on local topology, we will extend it to link prediction based on global topology in subsequent experiments and find similar conclusions.
	
	\section{Proposed methods}
	\label{sec:proposed}
	
	However, the situation of structural controllability is much more complicated. Regarding the core structure and edge structure of the network, structural controllability and structural predictability are obviously favored in completely opposite areas, which means that the two have different preferences in existing links. Specifically, on a real network, critical links may have weaker structural predictability, while the structural predictability of trivial links will be stronger. This conjecture based on theoretical analysis leads us to propose an index to quantify the relative structural predictability of critical links in the network, that is, the structural reciprocity index.

	\subsection{Structural reciprocity}
	The structural reciprocity index (SRI) can be defined as the probability that the predictability (see \textbf{Methods}) of a randomly selected critical link is less than one trivial link that is also selected randomly based on any particular prediction algorithm via leave-one-out process (Figure~\ref{illustration}c)
    \begin{equation}
    \nonumber
        SRI = \frac{n_1 + 0.5n_2}{n},
    \end{equation}
    where $n$ is the total number of sampling, $n_1$ and $n_2$ are respectively the number of sampling where the predictability of critical links are less or equal than that of trivial links. Obviously, the larger the $SRI$, the more evident the effect of structural controllability on predictability will be.\\
    	
    \subsection{Line graph}
    For the directed network $G(V,E)$ with $N$ nodes and $M$ edges, one can construct its corresponding undirected line graph $G^{L}(V',E')$ with $M$ nodes (Figure~\ref{illustration}b). Letting $V'=\{e|e\in E\}$, each node in $G^{L}$ represents a directed link in $G$. If two different links $e_i$ and $e_j$ share the same source or target node in $G$, set $e_i$ is adjacent to $e_j$ in $G^{L}$. The following centrality metrics \cite{newman2018networks} on $G^{L}$ are then adopted.
    \begin{itemize}
    \item[(i.)] \textbf{connectivity}. $c = \frac{1}{M}\sum_{i}k_i$, where $k_i$ is the degree of node $i$ and $M=|V'|$ is the number of nodes in $G^{L}$.
    \item[(ii.)] \textbf{closeness}. $C_i = M/\sum_{j\in V'}d_{ij}$, where $d_{ij}$ represents the shortest path length between node $i$ and $j$ in $G^{L}$.
    \item[(iii.)] \textbf{betweenness}. $B_i = \frac{2}{M(M-1)}\sum\limits_{s \neq t \in V'}\frac{g^i_{st}}{g_{st}}$, where $g^i_{st}$ represents the number of shortest paths from source node $s$ to target node $t$ through node $i$, and $g_{st}$ is the number of shortest path from node $s$ to node $t$ in $G^{L}$.
    \item[(iv.)] \textbf{k-shell decomposition}. $k_S$ is defined as the value of each node based on k-shell decomposition analysis~\cite{kitsak2010identification}. \\
    \end{itemize}
Connectivity describes the number of neighbors of a node in the line graph, which corresponds to the sum of the degree values of two adjacent nodes of this link in the original network, and intuitively reflects whether the link is located in the central area.
Any path in the line graph can be found in the original network. This makes it possible to describe the time required for any link in the original network to reach the remaining links based on the closeness centrality of the path length in the line graph. Similarly, the betweenness centrality based on the line graph is also used to characterize whether the corresponding node in the original network is located in the transportation hub of the reachable paths between any pair of links.
The k-core decomposition in the line graph can find a node cluster, which is the corresponding link cluster in the original network. This process can be understood as the k-core bond percolation~\cite{PhysRevLett.96.040601} in the original network.

\section{Experiments}
\subsection{Data description}
Here, we apply empirical analyses on eight following real networks.
\begin{itemize}
\item[(i.)]  \textbf{FWEW}~\cite{patricio2000network}: It contains 69 creatures living in Everglades Graminoids and 916 directed predatory relationships among creatures. In this network, each node represents a creature and a directed link $(i,j)$ represents creature $j$ is a predator of creature $i$.

\item[(ii.)]  \textbf{FWMW}~\cite{baird1998assessment}: FWMW is the food chain network of the mangrove estuary wet season, including 97 creatures and 1492 directed predatory relationships among creatures. In this network, each node represents a creature and a directed link $(i,j)$ represents creature $j$ is a predator of creature $i$.

\item[(iii.)] \textbf{FWFW}~\cite{ulanowicz1998network}: FWFW is the food web of the Florida Gulf during the rainy season, including 128 creatures and 2106 directed predatory relationships among creatures. In this network, each node represents a creature and a directed link $(i,j)$ represents creature $j$ is a predator of creature $i$.

\item[(iv.)] \textbf{Terrorist}: Terrorist represents ties among organizations, based on co-location of terrorists and a set of attacks they carried out. In this network, nodes are terrorist organizations and ties are co-location of terrorist attacks. This data can be accessed via \cite{pub2016}.

\item[(v.)] \textbf{Delicious}~\cite{lu2011leaders}: Delicious.com, the site not even allows users to rate and bookmark web pages, but also allows users to follow other users and collect their bookmarks and web pages. In this network, nodes represent users and a directed link $(i,j)$ means that user $i$ follows user $j$.

\item[(vi.)] \textbf{Kohonen}: Kohonen describes the directed citation relationship between papers about "Kohonen T". In this network, each node is a published paper and a directed link $(i,j)$ means that paper $i$ have cited paper $j$. This data can be accessed via~\cite{pub2006}.

\item[(vii.)] \textbf{SciMet}: SciMet is a citation among published papers in Scientometrics. In this network, each node is a published paper and a directed link $(i,j)$ means that paper $i$ have cited paper $j$. This data can be accessed via~\cite{pub2006}.

\item[(viii.)] \textbf{Guava}~\cite{vsubelj2012clustering}: Guava library dependencies is a network of software dependencies in Guava r07 core libraries. Nodes represent libraries, and a directed link indicates a dependency of one library on another.
\end{itemize}
\begin{table}[ht]
\centering
\caption{Basic topological statistics of eight real networks. $N$ and $M$ respectively represent the number of nodes and links, where ($\cdot$) is the number of critical links. $s=\frac{M}{N (N-1)}$ is the data sparsity.  $\langle k \rangle$ is the average degree. $cc$ is the clustering coefficient. $\langle d\rangle$ represents the average length of shortest path between all node pairs in weakly connected networks, in which `-' means that the network is not connected. $r$ is the assortativity coefficient. $SRI$ is the value of structural reciprocity index.}
\label{tab_real}
\tabcolsep 4pt
\begin{tabular}{lcccccccc}
\toprule
Datasets& \bfseries $N$ & \bfseries $M(\cdot)$ & $s$ & \bfseries $\langle k \rangle$ & $cc$& \bfseries $ \langle d\rangle $&$r$& $SRI$ \\
\hline
FWEW & 69 & 916 (1)  & 19.52\% & 13.28&0.3 & 1.89&	-0.4 & 0.81\\
FWMW & 97 & 1492 (1)  & 16.02\%  & 15.38&0.26 & 1.94&	-0.3& 0.45\\
FWFW & 128 & 2137 (3)  & 13.15\% & 16.7&0.17 & 1.95&	-0.23& 0.83\\
Terrorist &260 &571 (170)& 8.48\% &2.19 &0.31 &- &0.5& 0.74 \\
Delicious &300 &957 (30)& 10.67\% & 3.19&0.29& 0.53&	-0.45& 0.72 \\
Kohonen & 185 & 443 (36) & 13.01\% & 2.39&0.15& 0.11&	-0.17& 0.8\\
SciMet & 179 &  280 (28)  & 8.79\% & 1.56&0.06 & 0.02&	-0.3& 0.71\\
Guava&620 &1078 (51) & 2.81\% & 1.74& 0.12& -& 0.14& 0.61\\
\bottomrule
\end{tabular}
\end{table}

\begin{figure}[htbp]
    \centering
    \includegraphics[width=9cm,height=11.5cm]{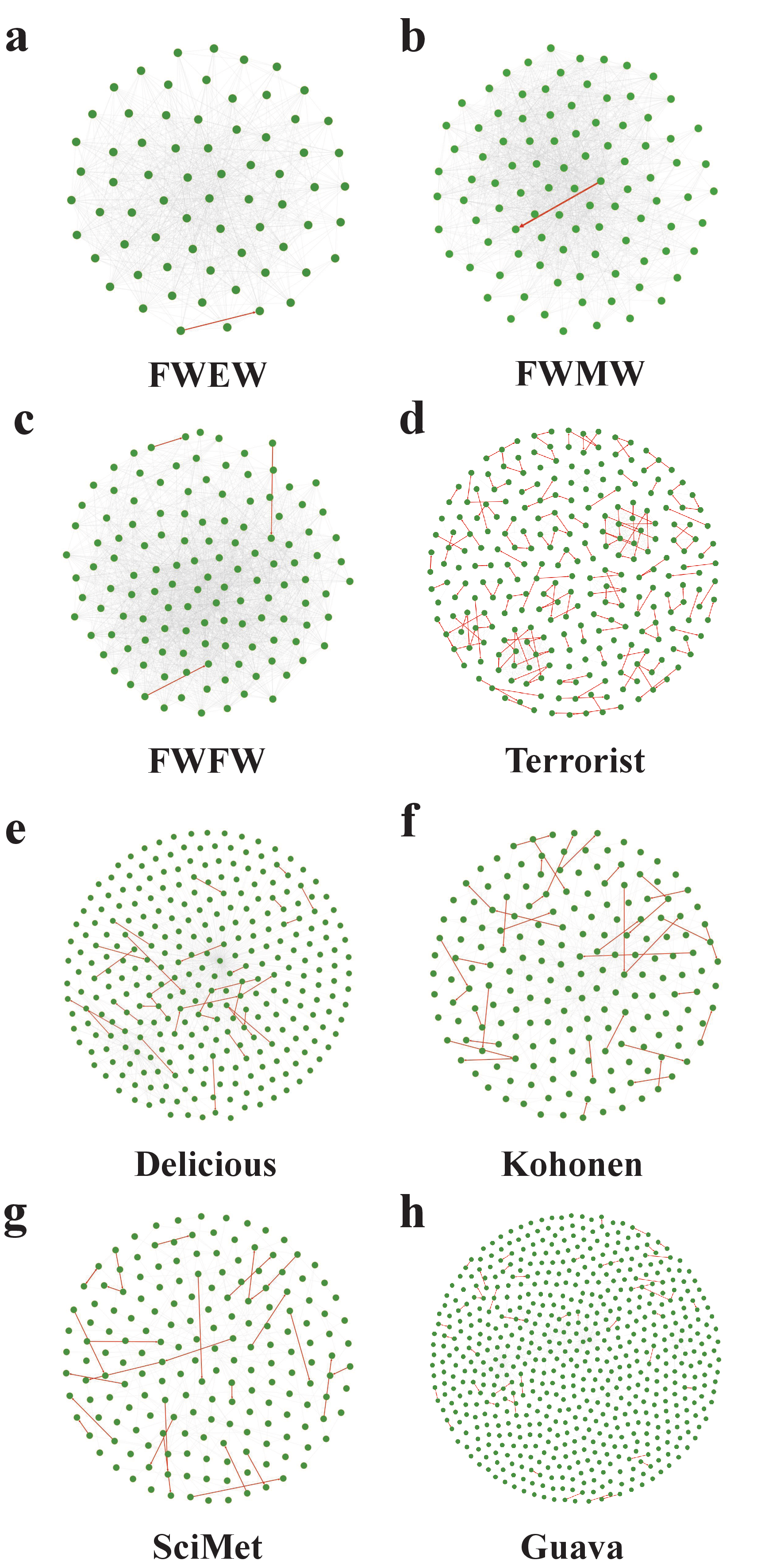}
    \caption{Visualization of the eight real networks. Red and gay lines represent critical and trivial links, respectively. Each arrow represent a directed link. It can be clearly seen that critical links only occupy a small fraction of every network.}
    \label{fig_data_visual}
\end{figure}

In this work, we construct two synthetic directed networks based on the classical models, described as
\begin{itemize}
\item[(i.)]\textbf{Directed Erdos-Renyi network} (ER): The classical ER network model~\cite{erdds1959random} is to describe a simple network generation process where every node pair connects with a given probability, and we randomly set the direction of links to generate a directed ER network.

\item[(ii.)] \textbf{Directed Barabasi-Albert network} (BA): The classical BA network model~\cite{barabasi1999emergence} characterizes the scale-free property of real complex networks, and we set the newly added nodes have directed outlinks to old ones according to the latter's indegree.
\end{itemize}
\begin{table}[htbp]
\footnotesize
\caption{Basic topological statistics of artificial networks. $M$ is the number of directed links, where ($\cdot$) in brackets means the average number of critical links. $\langle k \rangle$ is the average degree. $s=\frac{M}{N(N-1)}$ is the data sparsity. $cc$ is the clustering coefficient. $\langle d \rangle$ is the average length of shortest path between all node pairs in weakly connected networks, in which `-' means that this network is not connected. $r$ is the assortativity coefficient. $SRI$ is the value of structural reciprocity index. The number of nodes for all networks is set as $N=1000$. All the values are generated by averaging 10 independent realizations.}
\label{tab_artificial}
\tabcolsep 4pt
\begin{tabular}{lcccccccc}
\toprule
Network&$\langle k \rangle$&M($\cdot$)& $s$ &cc&r&$\langle d\rangle$&$SRI$\\
\hline
\multirow{10}*{\LARGE ER}&1&1002(321)& 0.1\%&0.0006&-0.0065&-&0.6\\
&2&1983(357)& 0.2\%&0.0019&-0.0032&-&0.61\\
&3&2978(272)& 0.3\%&0.003&-0.0044&5.6898&0.62\\
&4&4003(111)& 0.4\%&0.004&-0.0022&4.9114&0.69\\
&5&4997(42.3)& 0.5\%&0.0048&-0.0033&4.4103&0.73\\
&6&5989(19.7)& 0.6\%&0.0059&0.0004&4.0378&0.75\\
&7&7011(7.4)& 0.7\%&0.007&0&3.7575&0.76\\
&8&7958(3.6)& 0.8\%&0.0081&-0.0038&3.5613&0.78\\
&9&8997(1.7)& 0.9\%&0.009&-0.0001&3.3903&0.79\\
&10&9992(0.6)& 1\%&0.0101&-0.0015&3.2576&0.8\\
\hline
\hline
\multirow{10}*{\LARGE BA}&1&1195(163)& 0.12\%&0.0035&0.3467&0.0418&0.59\\
&2&2180(202)& 0.22\%&0.0756&0.3358&0.045&0.68\\
&3&3165(201)& 0.32\%&0.055&0.3135&0.0878&0.71\\
&4&4150(192)& 0.42\%&0.047&0.3042&0.1829&0.72\\
&5&5135(184)& 0.51\%&0.0413&0.2943&0.2997&0.74\\
&6&6120(165)& 0.61\%&0.0392&0.2852&0.3988&0.74\\
&7&7105(163)& 0.71\%&0.0383&0.277&0.487&0.76\\
&8&8090(155)& 0.81\%&0.0383&0.2659&0.5567&0.77\\
&9&9075(154)& 0.91\%&0.0386&0.2553&0.6298&0.77\\
&10&10060(138)& 1\%&0.0399&0.2463&0.6752&0.77\\%
\bottomrule
\end{tabular}
\end{table}

\subsection{Comparison Methods}
Each prediction algorithm is a derivative of the inherent evolutionary mechanism of the network which can score every potential link according to the observed network topology. Each can score every missing link, and we have selected several representative methods as benchmark algorithms to measure the link predictability. Given a simple directed network $G(V, E)$, we will introduce four benchmark methods as below.

\begin{itemize}
\item[(i.)] Common neighbor (CN) \cite{liben2007link}. Defined as the number of nodes connecting to both nodes $i$ and $j$.
\begin{equation}
\label{e10}
s_{ij}^{CN}= |\Gamma^{out}(i) \cap \Gamma^{in}(j)|,
\end{equation}
where $\Gamma^{out}(i)$ is the set of nodes that node $i$ links to, and $\Gamma^{in}(j)$ is the set of nodes that connect to node $j$.

\item[(ii.)]  Preferential attachment (PA) \cite{barabasi1999emergence}. Defined as the product of the respective degree of nodes $i$ and $j$.
\begin{equation}
\label{epa}
s_{ij}^{PA}= k^{out}_i k^{out}_j,
\end{equation}
where $k^{out}_i$ is the outdegree of node $i$.

\item[(iii.)]  Potential theory (Motif) \cite{zhang2013potential}. Defined as the number of Bi-fans that nodes $i$ and $j$ co-exist.
\begin{equation}
\label{emotif}
 s_{ij}^{motif}= |\{(u,v)\in E|u\in\Gamma^{in}(i),v\in\Gamma^{out}(j) \}|,
\end{equation}

\item[(iv.)]  LeaderRank (LR) \cite{lu2011leaders, fan2020finding} Defined as the sum of reaching probability from $i$ to $j$ via a random walk process by adding a ground node that connects to all the nodes of the network.
\begin{equation}
\label{e12}
s_{ij}^{LR}= \pi_{ij},
\end{equation}
where $\pi_{ij}$ is the $j$-th element at the stationary state of a random walk process $\pi^{t+1}_{i}=\sum_{j}\frac{a_{ij}}{k^{out}_j}\pi^{t}_j$, where $a_{ij}$ is the element of modified adjacent matrix of $G$ by adding the ground node, and $\pi^{0}$ is initially set as one for all nodes.
\end{itemize}

\subsection{Predictability of critical and trivial links}
\begin{figure}[!t]
    \centering
    \includegraphics[width=9.2cm,height=6.8cm]{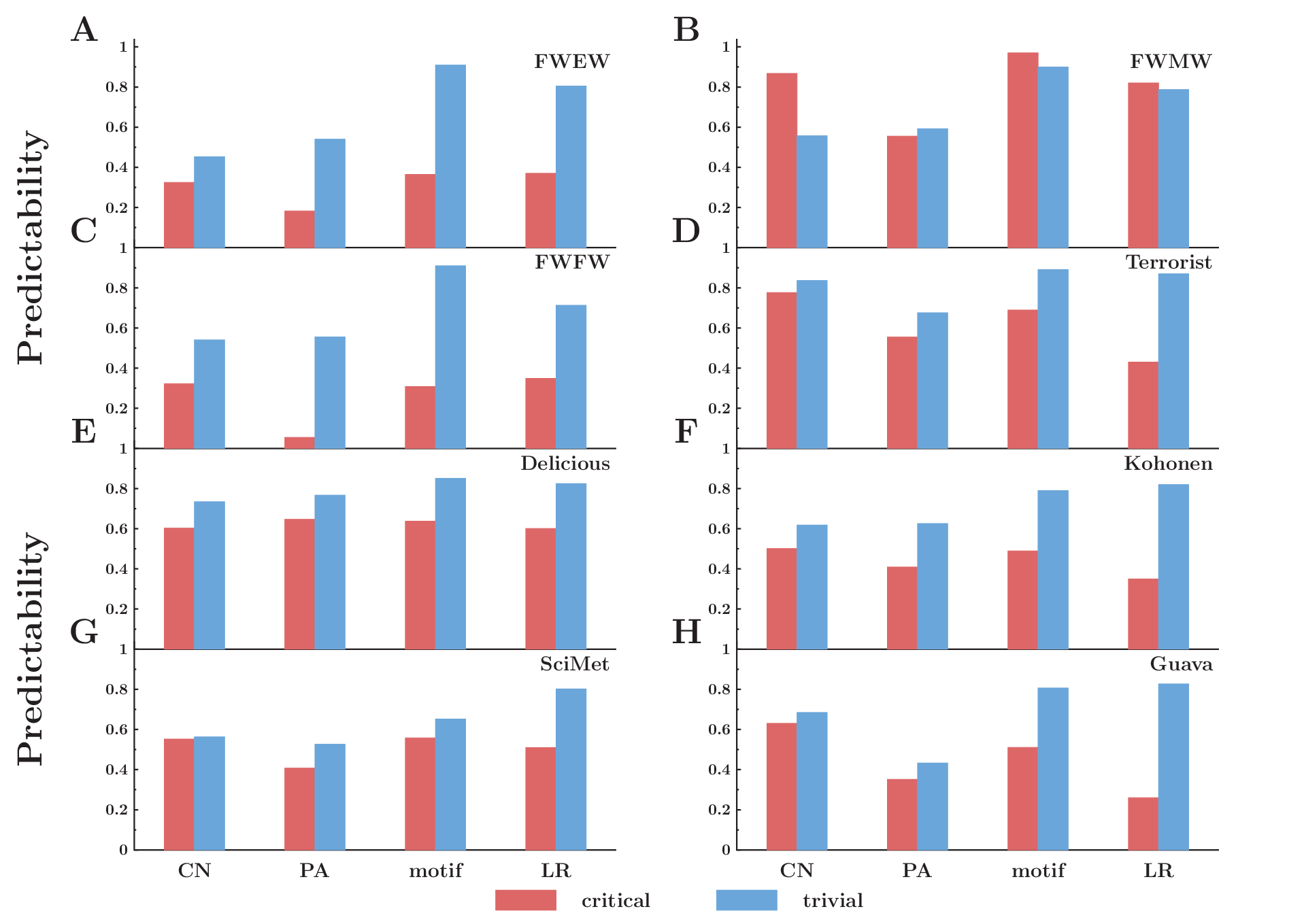}
    \caption{Predictability of critical (red) and trivial (blue) links in eight real networks based on four benchmark methods: (i) Common neighbor (CN)~\cite{liben2007link}, (ii) Preferential attachment (PA)~\cite{barabasi1999emergence}, (iii) Potential theory (Motif)~\cite{zhang2013potential}, and (iv) LeaderRank (LR)~\cite{lu2011leaders,fan2020finding}.}
    \label{fig1}
\end{figure}

Fig.~\ref{fig1} shows the experimental results of predictability for critical and trivial links in eight real networks based on four representative prediction algorithms. It can be seen that the predictability of critical links is significantly lower than that of trivial links in most datasets, except for \emph{FWMW} (Fig.~\ref{fig1}B), which represents an opposite case. Generally, critical links only occupy a small fraction of all of the network connections, leaving nearly all are the other trivial ones, e.g. more than 95\% in most real networks (TABLE~\ref{tab_real}). This will significantly affect the structural predictability. Taking the ER network as an example, the predictability of non-critical links is approximately equal to $0.5$ (Fig.~\ref{figS2}A-\ref{figS2}D), which is quite close to the case of a random guess. Comparatively, the predictability of critical links is even smaller than random selection, which makes it a rare circumstance to be predicted. If such links are divided into the probe set of a prediction task, it will inevitably make a tremendous influence on achieving the exact predictability of the target network. Furthermore, if the predictability of all of the links of a given network are randomly generated from an independent and identical distribution, the value of SRI should be around $0.5$. Therefore, the degree to which of SRI exceeds $0.5$ indicates how significant the effect of structural reversibility is. It shows that the $SRI$ values of all the eight observed real datasets and two synthetic networks (see TABLE~\ref{tab_real}, TABLE~\ref{tab_artificial} and Fig.~\ref{figS2}) are larger than $0.5$ expect \emph{FWMW}, which further confirms that the correlation between structural controllability and predictability may universally occur in complex networks.\\

\subsection{Effect of fraction of critical links on structural predictability}
\begin{figure}[!t]
    \centering
     \includegraphics[width=8.5cm,height=8.8cm]{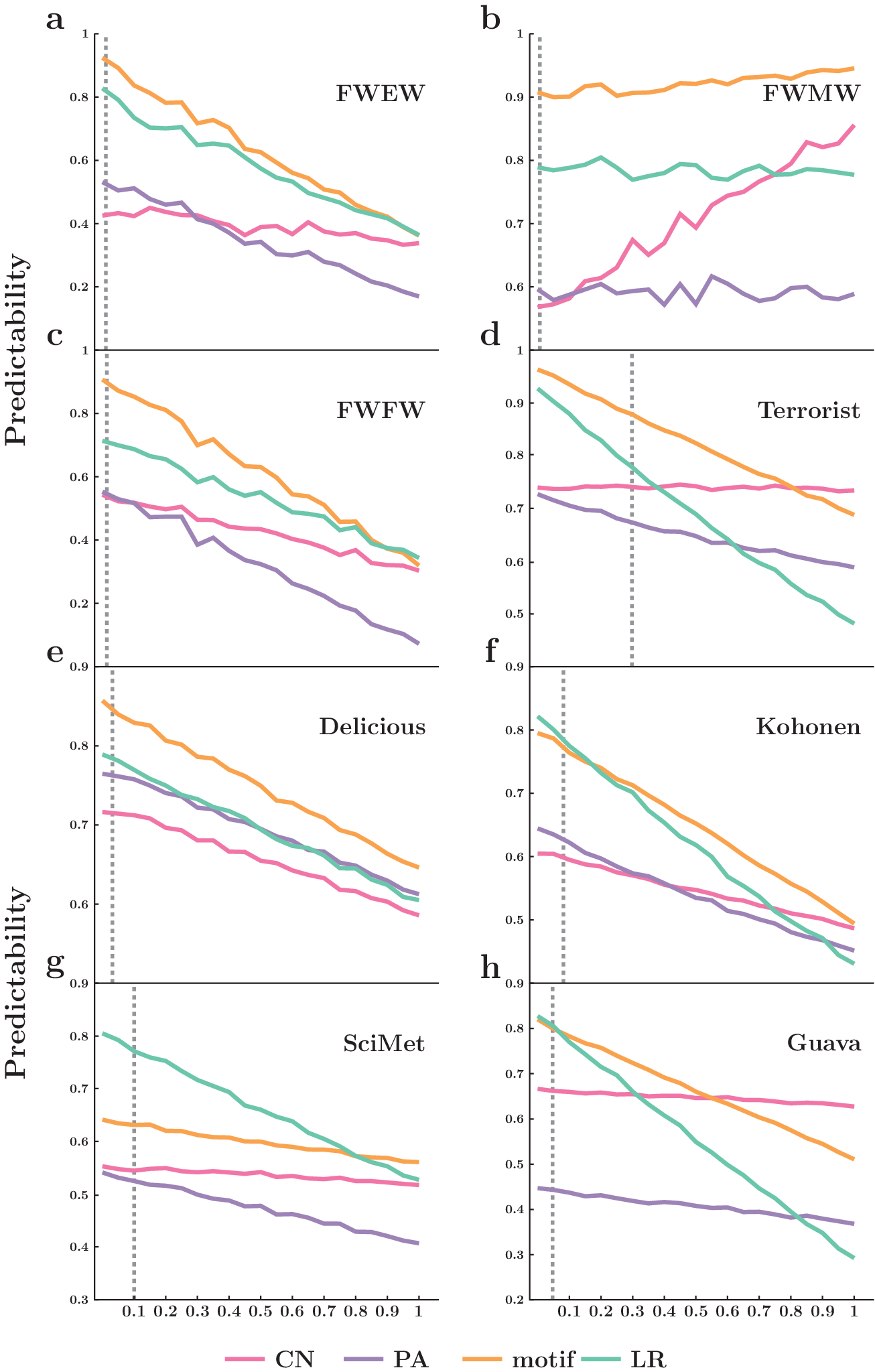}
   \caption{Structural predictability as the function of the fraction of critical links in the probe set for the eight real networks and four representative prediction algorithms.
     Gray dashed line represents the ratio of critical links in their respective original networks. The selection radio of the test set is 0.1.}
    \label{criticalratio}
\end{figure}

For traditional prediction algorithms, an arbitrary number of links will be randomly drawn into the probe set to evaluate their performance and robustness, regardless of the link properties. If the random selection is evenly drawn from a uniform distribution, the ratio of critical links in the original network will be equivalent to that in the probe set. Therefore, we subsequently check the impact of the ratio of critical links on the predictability. Fig.~\ref{criticalratio} shows that structural predictability will generally decrease along with the increase in the number of critical links in the probe set for all of the datasets, except \emph{FWMW}. Similar results can also be found on synthetic networks (Fig.~\ref{figS3}). This reveals the fact that one may obtain an incorrect evaluation result for the algorithms if disproportionate critical links are added to the probe set, suggesting that choosing testing links should be done very carefully to avoid the overfitting/underfitting problem of optimization~\cite{hawkins2004problem,srivastava2014dropout}.

Note that the ratios of critical links in the first three datasets (\emph{FWEW},\emph{FWMW} and \emph{FWFW}), approximately 1\%, are surprisingly lower than those of other five networks, making them all have very high predictability (gray dashed line in Fig.~\ref{criticalratio}). Comparatively, \emph{Terrorist} and \emph{SciMet} have larger numbers of critical links (approximately 29.8\% and 10\%, respectively), and achieve relatively lower predictability. Note that these two networks are also very sparse (TABLE.~\ref{tab_real}), which suggests that they are more difficult to control along with the large fraction of critical links, which may mainly locate at the periphery of  the network. On the contrary, although the sparsity of \emph{Guava} is the lowest, its ratio of critical links is only approximately 4.7\%, which suggests that it is not difficult to control. It additionally demonstrates that a more easily controllable network is expected to have higher structural predictability, and vice versa.\\

\subsection{Link Centrality on Line Graph}
To further investigate the effect of different type of links on the structural predictability, the original network was converted to a line graph that treats links as nodes (Fig.~\ref{illustration}b) \cite{Shah1995Conceptual,evans2009line}. We are then able to adopt various node centrality indices in the corresponding line graph to describe the topological centrality of links which is relatively difficult to quantify in the original network. Table~\ref{tab_line_graph} shows the differences between the critical and trivial links for the observed real networks. It can be seen that all of the centrality measures of critical links are significantly lower than those of the trivial links in all of the real networks except for \emph{FWMW}. As discussed in Fig.~\ref{illustration}, the predictability of links is highly dependent on their respective location in the networks, resulting in the fact that links in the central area are more predictable than those at the periphery of the network. Therefore, critical links based on the structural controllability of the entire network are more likely to be located at the periphery than that of trivial links, since the node centrality obtained from the line graph reflects the topological properties of corresponding links.

\begin{figure}[!t]
    \centering
     \includegraphics[width=9.0cm,height=8.5cm]{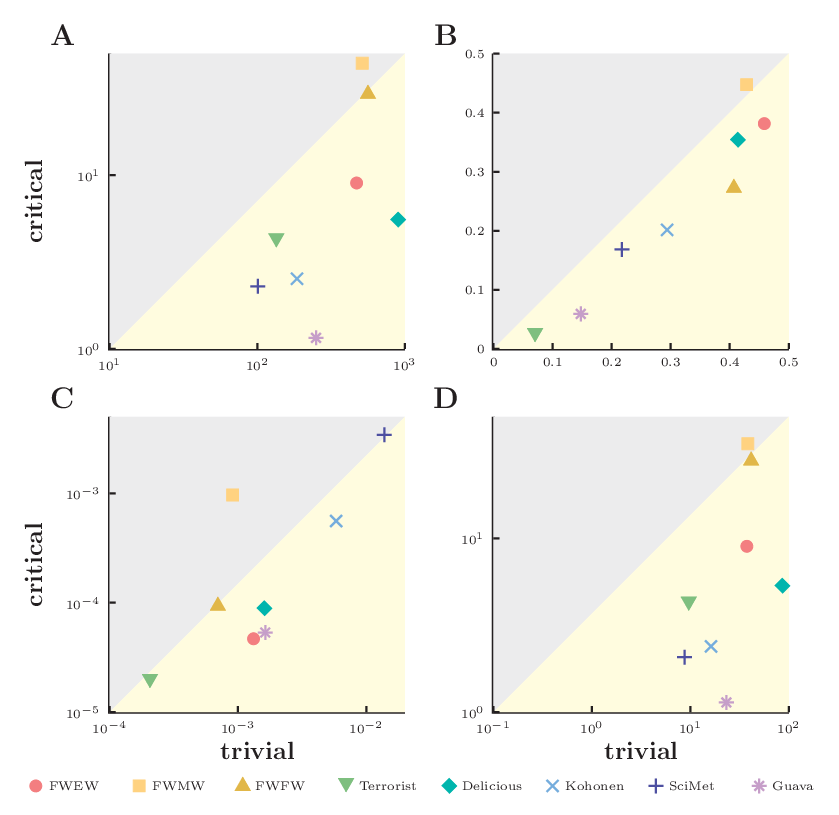}
    \caption{Link centrality of critical links versus trivial links for eight real networks. Link centralities are calculated in the corresponding line graphs, including (a) degree, (b) closeness, (c) betweenness, and (d) k-shell decomposition. }
    \label{fig7}
\end{figure}

It is worth noting that the results for \emph{FWMW} show an exceptional case in which almost all the link centrality properties are equivalent for critical and trivial links. One natural explanation is that, different from other datasets, the majority of critical links may be situated in the center rather than at the periphery in \emph{FWMW} (Fig.~\ref{fig_data_visual}), as illustrated in Fig.~\ref{illustration}. In Fig.~\ref{fig7}, the relationship between critical and trivial links for the examined link centrality measures is also plotted. It can be seen that each subfigure of Fig.~\ref{fig7} is equally divided into two parts by a diagonal line, and the gray area represents the link centrality effect of critical links being stronger than that of trivial links, while the light-yellow area shows the converse case. As a consequence, the link centrality results of \emph{FWMW} mainly fall in the gray area, suggesting that its critical links have a significant impact on improving the corresponding structural predictability, as shown in Fig.~\ref{fig1}B. \\

\begin{figure}[tbp]
    \centering
    \includegraphics[width=9cm,height=8.5cm]{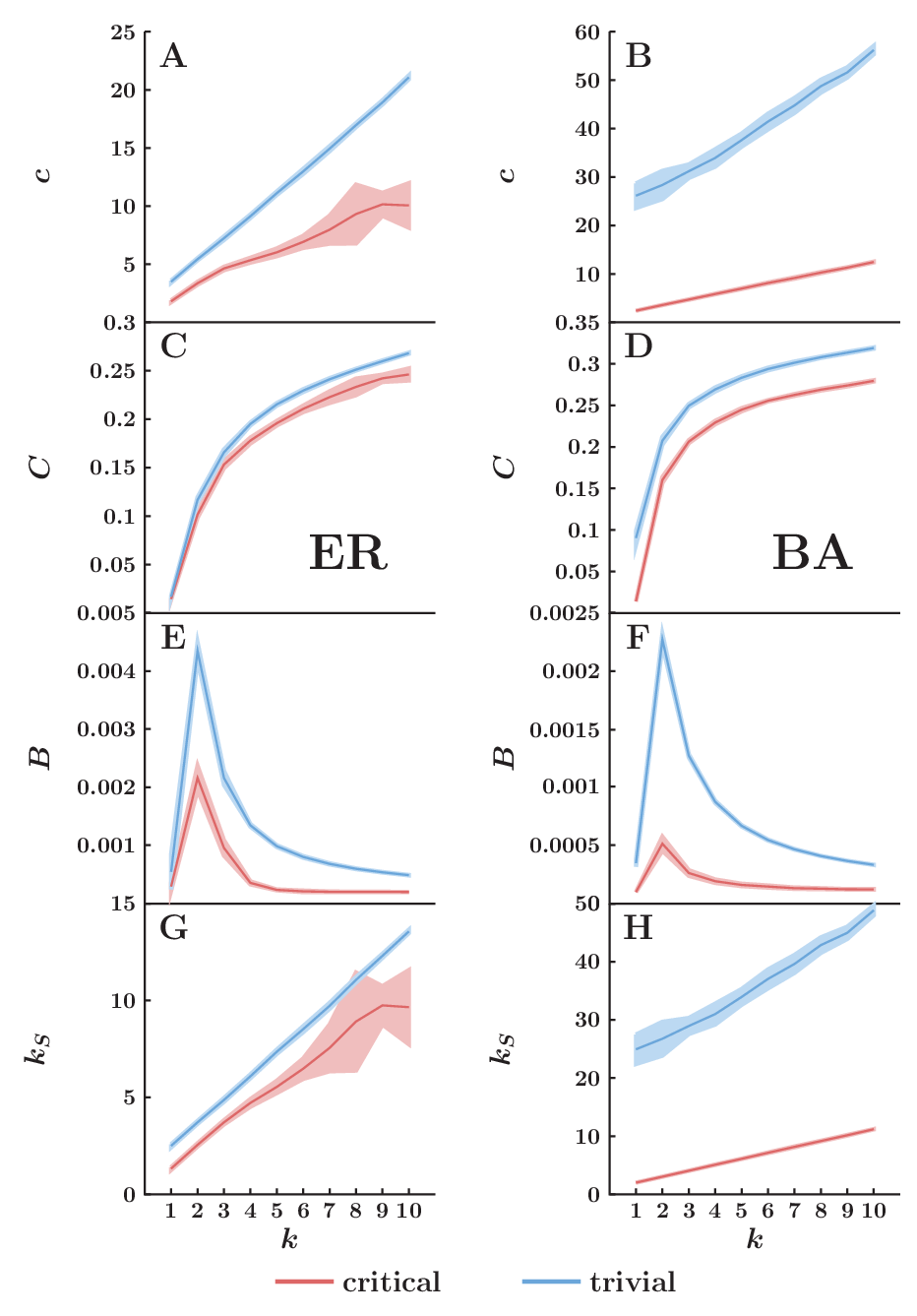}
    \caption{Topology centrality of critical (red) and trivial links (blue) as the function of data density for synthetic networks based on the line graph transformation.
    (a, c, e, g) and (b, d, f, h) respectively show the topology features based on ER and BA networks. The network size is set as $N=1000$. Light shadows represent the error bar.}
    \label{figS4}
\end{figure}

\subsection{Synthetic networks}
\begin{figure*}[!htbp]
    \centering
    \includegraphics[width=\linewidth]{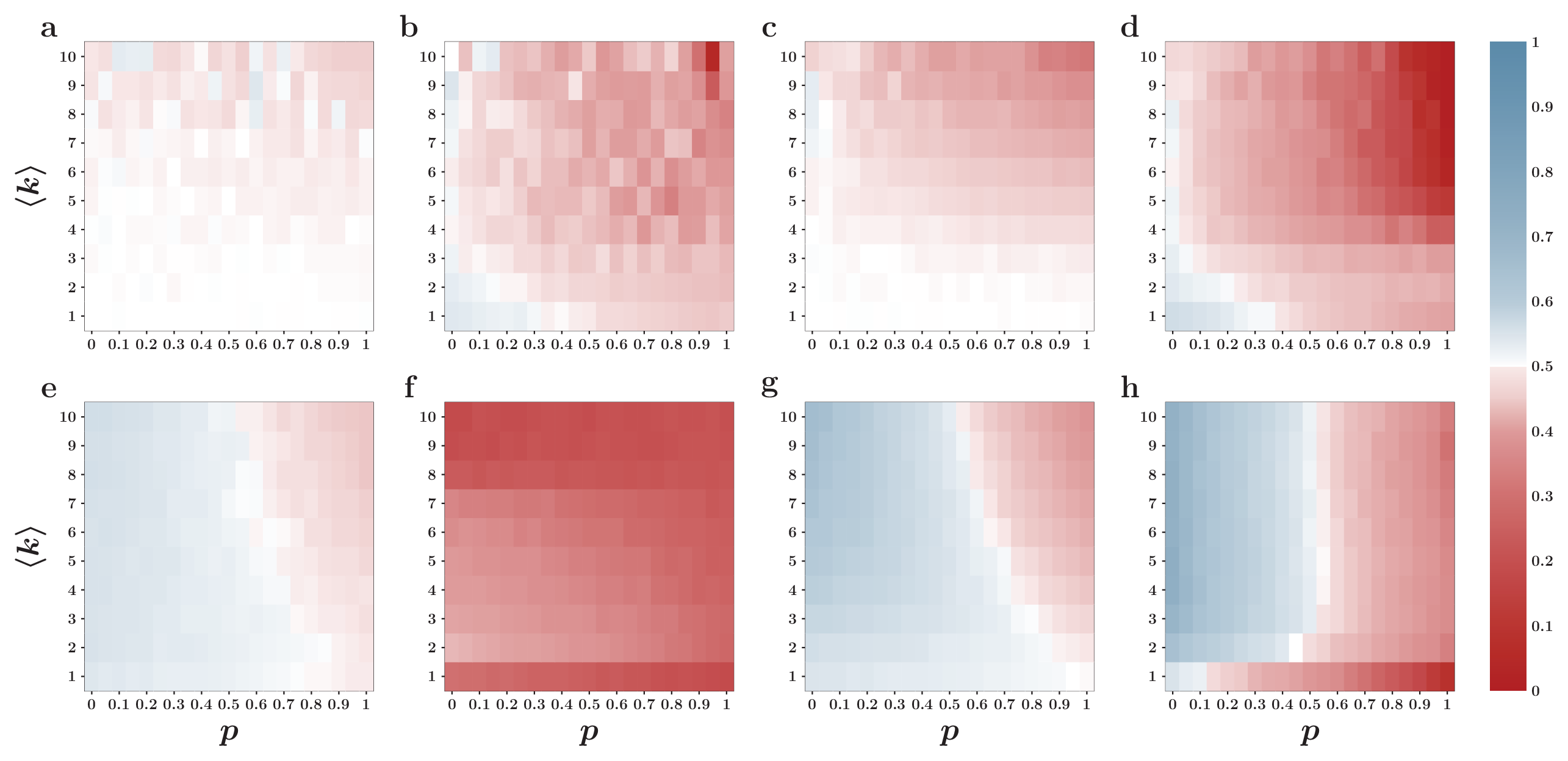}
    \caption{Structural predictability as the function of network density ($\langle k \rangle$) and the fraction of critical links in the probe set ($p$) in ER (A-D) and BA (E-H) networks based on four representative prediction methods:  (A,E) CN, (B,F) PA, (C,G) Motif, and (D,H) LR. The network size $N$ is set as 1000, and each prediction is obtained by averaging over 10 independent selections. It shows that the more critical links in the probe set, the less predictable of the network.  }
    \label{fig_artificial_auc}
\end{figure*}
To further investigate the relationship between structural controllability and predictability, similar experiments were also performed on two synthetic datasets: Barabasi-Albert (BA) ~\cite{barabasi1999emergence} and Erdos-Renyi (ER)~\cite{erdds1959random} networks. Fig.~\ref{fig_artificial_auc} shows that the predictability of critical links is lower than that of others on both networks with virous densities. In particular, the difference between the two types of links increases along with network density as large-degree nodes generally have more information, e.g., more common neighbors, and hence can be more easily to be predicted. This phenomenon is largely due to the fact that critical links are more likely to be situated at the periphery of both ER and BA networks. As a consequence, critical links as necessary hubs from driver nodes to others will only have little information and will be less likely to be identified.
\begin{table}[htbp]
		\centering
		\caption{Topology centrality statistics of critical and trivial links in eight real networks based on the line graph transformation. $c$ represents connectivity, $C$ is closeness, $B$ is betweenness, and $k_{S}$ is average value based on k-shell decomposition.}
		\begin{tabular}{p{5em}ccccc}
			\hline
			Datasets&Type&$c$&$C$&$B$&$k_{S}$ \\
			\hline
			\multirow{2}*{FWEW}&critical&9.01&0.382&0.00005&9\\
	&trivial&47.09&0.459&0.00132&37.5\\
\hline
&critical&44.03&0.447&0.00097&35\\
\multirow{-2}*{FWMW}&trivial&51.44&0.428&0.00091&38.36\\
\hline
\multirow{2}*{FWFW}&critical&29.35&0.273&0.00009&28\\
	&trivial&55.99&0.407&0.0007&41.26\\
\hline
\multirow{2}*{Terrorist}&critical&4.34&0.027&0.00002&4.33\\
	&trivial&13.42&0.07&0.0002&9.58\\
\hline
\multirow{2}*{Delicious}&critical&5.54&0.354&0.00009&5.33\\
	&trivial&89.85&0.414&0.0016&85.95\\
\hline
\multirow{2}*{Kohonen}&critical&2.53&0.2&0.00056&2.39\\
	&trivial&18.58&0.29&0.0058&16.24\\
\hline
\multirow{2}*{SciMet}&critical&2.29&0.169&0.0034&2.07\\
	&trivial&10.09&0.217&0.0138&8.72\\
\hline
\multirow{2}*{Guava}&critical&1.16&0.059&0.00005&1.13\\
&trivial&25.01&0.148&0.0016&23.23\\
			\hline
		\end{tabular}%
		\label{tab_line_graph}%
	\end{table}%

\begin{figure}[htbp]
    \centering
    \includegraphics[width=\linewidth]{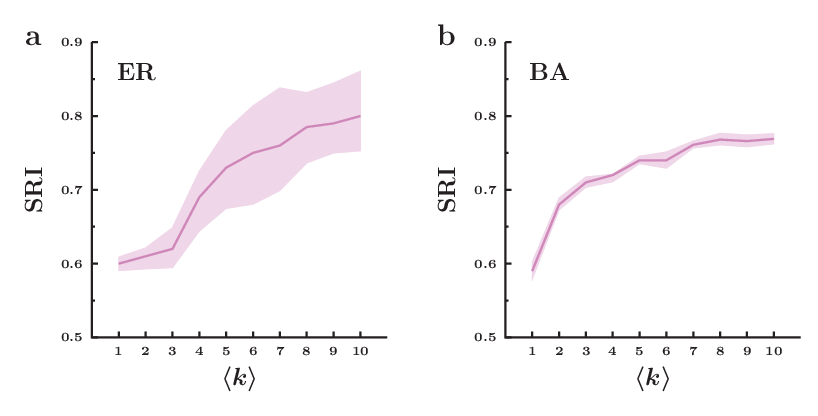}
    \caption{$SRI$ values as the function network density ($\langle k \rangle$) for (a) ER and (b) BA networks. Light shadows represent the error bar.}
    \label{figS3}
\end{figure}

\begin{figure*}[htbp]
    \centering
    \includegraphics[width=14.5cm,height=8.2cm]{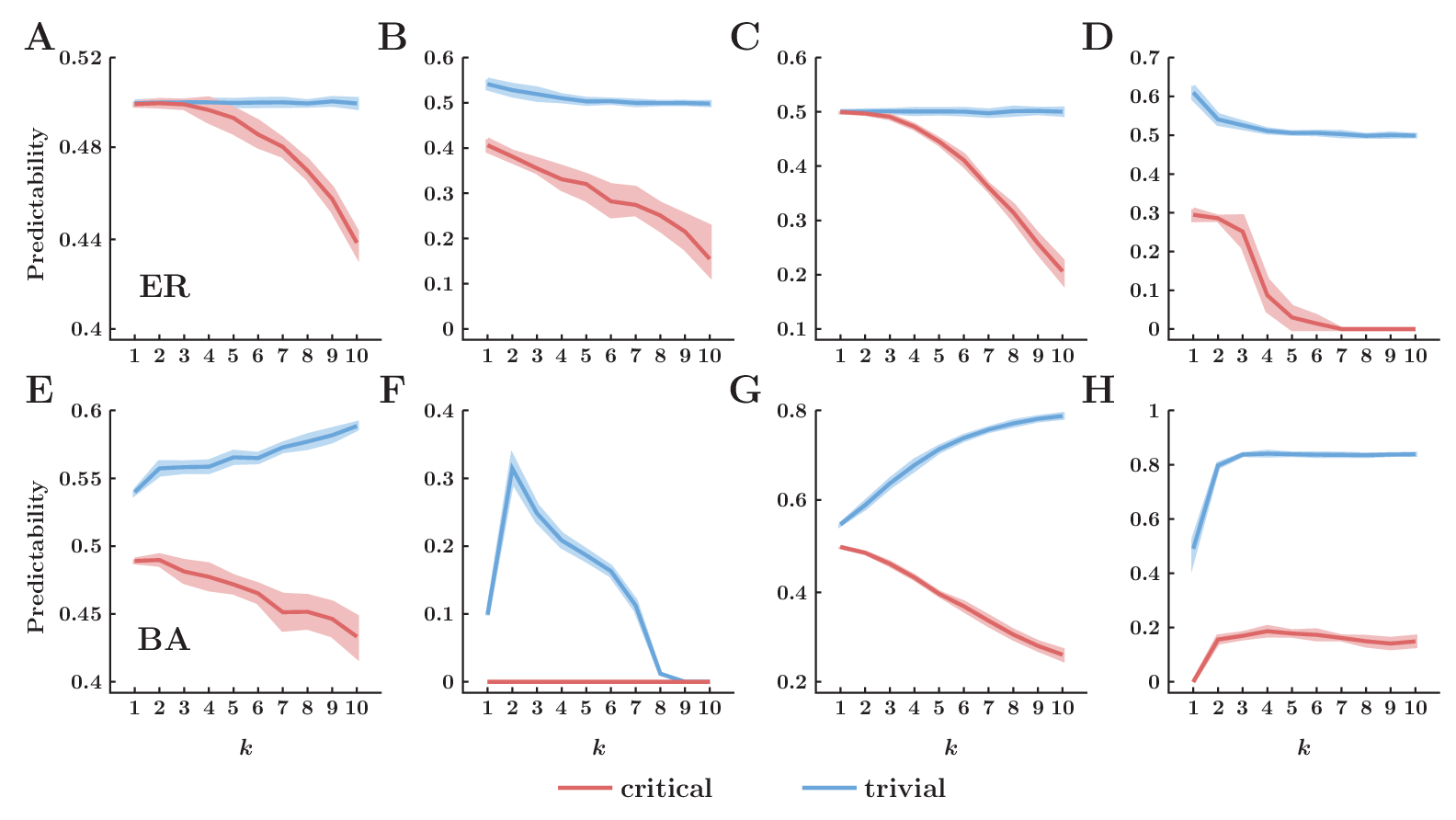}
    \caption{Predictability as the function of the network density for critical (red) and trivial (blue) links in two artificial networks: (A-D) ER network, and (E-H) BA network. Each sub-figure represents the results of respective benchmark methods: (A,E) CN, (B,F) PA, (C,G) Motif, and (D,F) LR. Network size $N$ is set as 1000. Each prediction is obtained by averaging over 10 independent selections, and light shadows represent the error bar.}
    \label{figS2}
\end{figure*}

We also observe the effect of critical links on predictability for various data sparsities on both networks (Fig.~\ref{figS2}). It can be seen that the value of \emph{SRI} increases as the data become denser for both networks, suggesting that the influence of structural reciprocity is more significant for dense data, which is also validated in Fig.~\ref{figS3}). Owing to the apparent structural difference, the ER network tends to be split into isolated communities when the network is too sparse, resulting in more critical links for sparse data and more non-critical links for a high-density network. On the contrary, although the ratio of critical links decreases slightly with increasing data density for BA network the value of $SRI$ still increases along with the increasement of data density. In the BA network, nodes are inclined to connect with large-degree ones, leading to a large number of small-degree driver nodes, thus making the network difficult to control. As the consequence, the critical links that attach those small-degree ones to central nodes are relatively easily predicted.

In addition, the link centrality properties for both synthetic networks based on the corresponding line graphs are shown comprehensively in Fig.~\ref{figS4}.
Similar to the experimental results of real networks, it can be seen that the values of critical links are generally smaller than trivial links in all centrality metrics. As the closeness index represents the marginal characteristic (Fig.~\ref{figS4}C and Fig.~\ref{figS4}D) of critical links from the perspective of the dynamics process, it suggests that critical links lie not in the central area of the two synthetic networks, but they locate very far from the core region, and difference between the two links is significantly larger on BA networks. This phenomenon is additionally demonstrated by the k-shell decomposition analysis (Fig.~\ref{figS4}G and Fig.~\ref{figS4}H) where the vital nodes in the transformed line graph shall be inclined to be at the core region. Results from betweenness index (Fig.~\ref{figS4}E and Fig.~\ref{figS4}F) further suggest that the small fraction of critical links makes little contribution to the network connectivity for both ER and BA graphs.
Besides, the non-monotonic curves in Fig.~\ref{figS4}E and Fig.~\ref{figS4}F originate from the fact that most of the node pairs in the network are unreachable when $k=1$.
\\

\section{Conclusions}
The present work demonstrates that critical links play an important role in maintaining the underlying relationship between structural controllability and predictability. This so-called structural reciprocity phenomenon widely exists in natural and social directed networks. The existence of critical links not only affect the number of driver nodes that further affect the difficulty level of network controllability, but they can also influence the upper bound of network predictability. Experimental results from representative real and synthetic networks show that both the fraction and position of critical links have significant impact on the prediction of target graphs. In most cases, critical links tend to be situated on the periphery of networks and have little information, and hence are less likely to be identified. However, under some circumstances, critical links may also locate at the center of networks to carry rich information to be easily restored. In addition, critical links only account for a small proportion in most controllable networks and will not significantly influence their predictability. Comparatively, they will considerably reduce the predictability for networks that are difficult to control due to the fact that the large number of critical links have rich and unrecoverable information. The importance of critical links reveals the ``stubborn links''\cite{altafini2014predictable} in sparse networks. Not only are they highly vulnerable to cyber attacks, but they are also extremely difficult to recover after being perturbed or destroyed.

As structural controllability points out the shortcomings of traditional link prediction methods, combining the two to propose new and improved methods is also a potential focus of future work. Take the common neighbor method as an example. Since structural controllability gives each link a "critical" or "trivial" attribute, a natural idea is that we can assign different weights to links according to their attributes. Here we give a brief example, combining structural controllability and common neighbor algorithm
\begin{align}
    s_{ij}^{CN} = \sum\limits_{k \in \Gamma_{ij}} w_{ik} + w_{jk} \nonumber
\end{align}
in which $\Gamma_{ij}=\Gamma^{out}(i) \cap \Gamma^{in}(j)$. $w_{ik}=1$ if $(i,k)$ is a trivial link in the training set; otherwise $w_{ik}=0.5$. Another idea is that if the newly added link is a critical link, the possibility of joining the network will be greatly reduced.
\begin{align}
   s_{ij}^{CN} = w'_{ij} |\Gamma_{ij}|, \nonumber
\end{align}
in which $w'_{ij}=1$ if $(i,j)$ is a trivial link after $(i,j)$ is added into training set; otherwise $w_{ik}=0.5$.

In conclusion, this work provides another perspective on network controllability and predictability, rooted in features of topological structure, leading us to understand how different links affect network functions at different scales. It enables us to evaluate the performance of prediction algorithms from the microscopic point of view, and to test which types of links are vital in social interactive systems. Most importantly, it also points out that a promising direction for prediction tasks is to focus on maximizing the accuracy of special links, rather than making efforts to forcibly enhance the overall predictability of all links.\\

\section*{Acknowledgments}
This work is supported by the National Natural Science Foundation of China (Grant Nos. 61873080, 92146001, 11971270, 61673151 and 11631014), the  Natural Science Foundation of Zhejiang Province (Grant No. LR18A050001), Natural Science Foundation of Shandong Province (Grant Nos. ZR2018MA001 and ZR2019MA047), and the Major Project of The National Social Science Fund of China (Grant No. 19ZDA324).

\ifCLASSOPTIONcaptionsoff
\newpage
\fi

\bibliographystyle{ieeetr}
	
\bibliography{references}

\end{document}